\documentclass[useAMS,usenatbib,usegraphicx]{mn2e}

\usepackage[totalwidth=480pt,totalheight=680pt,letterpaper]{geometry}

\usepackage{rotating}
\usepackage{lscape}
\usepackage{color}
\usepackage{amsmath}
\usepackage{hyperref}

\newcommand{\be}{\begin{equation}}
\newcommand{\ee}{\end{equation}}
\newcommand{\bea}{\begin{eqnarray}}
\newcommand{\eea}{\end{eqnarray}}

\def\se#1{Section~\ref{sec:#1}}

\def\Fig#1{Figure~\ref{fig:#1}}

\def\ifm#1{\relax\ifmmode#1\else$\mathsurround=0pt #1$\fi}
\def\kms{\ifmmode\,{\rm km}\,{\rm s}^{-1}\else km$\,$s$^{-1}$\fi}

\def\msun{\,{\rm M_{\odot}}}
\def\Msun{\,{\rm M_{\odot}}}

\def\ltsima{$\; \buildrel < \over \sim \;$}
\def\simlt{\lower.5ex\hbox{\ltsima}}
\def\gtsima{$\; \buildrel > \over \sim \;$}
\def\simgt{\lower.5ex\hbox{\gtsima}}

\def\r200{r_{200}}
\def\m200{m_{200}}
\def\V200{V_{200}}

\def\rmax{r_{\rm max}}
\def\Vmax{V_{\rm max}}
\def\Vmaxmax{V_{\rm pmax}}

\begin{document}


\title[Shapes and orientations of satellite haloes]{Galactic Tides and the Shape and Orientation of Dwarf Galaxy Satellites}
\author[C. Barber
et al.]  {Christopher Barber$^{1}$, Else Starkenburg$^{1,2}$, Julio F. Navarro$^{1,3}$, Alan W. McConnachie$^{4}$\\
$^{1}$Dept. of Physics and Astronomy, University of Victoria, P.O. Box 1700, STN CSC, Victoria BC V8W 3P6, Canada \\
$^{2}$CIFAR Global Scholar and CITA National Fellow\\
$^{3}$CIFAR Senior Fellow \\
$^{4}$NRC Herzberg Institute of Astrophysics, 5071 West Saanich Road, Victoria,
British Columbia, Canada, V9E 2E7}



\pagerange{\pageref{firstpage}--\pageref{lastpage}} \pubyear{2014}

\maketitle

\label{firstpage}

\begin{abstract}
  We use cosmological $N$-body simulations from the Aquarius Project to study the tidal effects of a dark matter halo on the shape and orientation of its substructure. Although tides are often assumed to enhance asphericity and to stretch subhaloes tangentially, these effects are short lived: as in earlier work, we find that subhaloes affected by tides become substantially more spherical and show a strong radial alignment toward the centre of the host halo. These results, combined with a semi-analytic model of galaxy formation, may be used to assess the effect of Galactic tides on the observed population of dwarf spheroidal (dSph) satellites of the Milky Way and Andromeda galaxies. If, as the model suggests, the relatively low dark matter content of luminous dSphs such as Fornax and Leo I is due to tidal stripping, then their gravitational potential must be substantially more spherical than that of more heavily dark matter-dominated systems such as Draco or Carina.  The model also predicts a tidally-induced statistical excess of satellites whose major axis aligns with the direction to the central galaxy. We find tantalizing evidence of this in the M31 satellite population, which suggests that tides may have played an important role in its evolution. \end{abstract}


\section{Introduction}
\label{SecIntro}

The dwarf spheroidal (dSph) galaxies around the Milky Way (MW) are important testbeds of current models of galaxy formation and cosmology. Their mass-to-light ratios vary from $\sim 10$ to $\sim 1000$ \citep{Mateo1998, Gilmore2007, Walker2013}, implying that the gravitational potential felt by stars in many of these galaxies---and thus their internal dynamics---is dominated by dark, rather than baryonic, matter. Line-of-sight (LOS) velocities of individual stars in dSphs can therefore be used to place useful constraints on the structure of their dark matter haloes \citep[e.g.,][]{Kleyna2002, Walker2007, Mateo2008, Battaglia2013}.

The interpretation of radial velocity data, however, is sensitive to assumptions, such as spherical symmetry, that are often used but difficult to verify or justify. \citet{Hayashi&Chiba2012}, for example, have argued that the enclosed mass within $300$ pc inferred from LOS velocity data can vary by up to a factor of ten when the assumption of spherical symmetry is lifted. Furthermore, the distribution of stars in many dSphs is clearly non-spherical \citep[e.g.,][]{Irwin&Hatzidimitriou1995, Martin2008}, and they are expected to inhabit dark matter haloes whose shapes are predicted to be strongly triaxial, at least in the current paradigm for structure formation, $\Lambda$CDM \citep{Frenk1988,Dubinski91,Jing&Suto2002, Hayashi2007, VeraCiro2011, Allgood2006, Kuhlen2007, Schneider2012}. Although substructure haloes have been found to be less triaxial than field haloes of comparable mass \citep{Kuhlen2007,Pereira2008,VeraCiro2014}, these departures from sphericity may still substantially affect the model results. Constraining the shape of the gravitational potential of dSph haloes thus seems crucial to furthering our understanding of these faint systems.

If dSph potentials are indeed triaxial, LOS velocity studies will also be sensitive to the orientation of dSph haloes with respect to the line of sight. While a random distribution might be naively expected, numerical simulations have consistently indicated that substructure haloes (or ``subhaloes'', the presumed hosts of dSph satellites) align so that their major axes tend to point to the centre of the main halo \citep[e.g.,][]{Kuhlen2007, Faltenbacher2008, Knebe2008, Knebe2010, Pereira2008, Pereira2010, VeraCiro2014}.  

A number of scenarios have been proposed in the literature to explain this alignment. A primordial origin in which haloes are tidally torqued by the surrounding large-scale tidal field seems implausible, for in that case the alignment would be strongest in the outskirts of the main system, something that is not observed. More recently, \citet{Pereira2005} have proposed that the tidal interactions between satellites and their host systems may be the culprit. The exact mechanism, however, is still debated. Some studies \citep[e.g.,][]{Pereira2008, Pereira2010, Knebe2010} have suggested that tidal torquing by the host halo is the main cause, while others \citep[e.g.,][]{Kuhlen2007} have proposed that the alignment may be due to stretching along the direction of (highly eccentric) orbits as a result of tidal stripping.

Interestingly, radial alignments have also been reported in galaxy clusters and groups \citep{Hawley&Peebles1975, Djorgovski1983, Pereira2005, Agustsson2006, Faltenbacher2007}, although it is still unclear whether there is quantitative agreement between the magnitude of the predicted and observed effects \citep[e.g.,][]{Bernstein&Norberg2002, Adami2009, Hung2012, Schneider2013, Sifon2014}.  Such intrinsic alignments may also affect the calibration of weak lensing surveys, which often assume that galaxies are randomly oriented in space \citep{Smith2001, McKay2001}. For example, \citet{Schneider2013} estimate that intrinsic alignments between galaxy group members may contribute a systematic uncertainty of up to 20 per cent in the mean differential projected surface mass density inferred for those systems. A thorough understanding of the physical origin of the alignment seems needed in order to assess the importance of this bias in the interpretation of observations.

We address these issues here using cosmological simulations of the formation of Milky Way-sized dark matter haloes from the Aquarius Project \citep{Springel2008}. Our study builds upon the earlier studies cited above but extends them in several respects. Rather than analyzing the properties of the subhalo population at large, we focus on those subhaloes that, according to a semi-analytic model of galaxy formation, are likely hosts of the ``classical'' (i.e., $M_V<-8$) dSph population of the Milky Way. This specific focus is important since the majority of subhaloes are expected to be dark and might therefore have properties different from those that harbour luminous satellites. Our study also differs from earlier work because we measure the shape of the gravitational potential (rather than that of the mass distribution) at radii comparable to the typical half-light radii of dSphs, where stars are found. This allows our results to be directly applied to dynamical models of observed dSphs with little extrapolation.

One shortcoming of our analysis is that it is based on dark matter-only simulations, and thus neglects the effects that baryons may have on the mass profile and the shape of galaxies and their haloes \citep[e.g.,][]{Navarro1996b, Read&Gilmore2005, Mashchenko2008, Knebe2010, Kazantzidis2010, Abadi2010, Penarrubia2012, Pontzen&Governato2012, Zolotov2012, Garrison-Kimmel2013, Arraki2014, Brooks2014, DiCintio2014}. Although we expect that these effects should be modest in dark matter-dominated systems such as dSphs, this is clearly an issue that should be revisited once realistic hydrodynamical simulations of the formation of faint satellite systems in a cosmological context become possible. 

This paper is organised as follows. In \se{sims} we describe the numerical simulations and semi-analytic model used to identify luminous satellites around MW-sized haloes. Our potential-fitting method is described and tested for numerical convergence in \se{analysis}. In \se{shapes} we show results from our shape measurements, while we discuss the radial alignment of dSph haloes and compare our alignment results with observational data for M31 satellites in \se{orientation}. We conclude with a brief summary in \se{conclusions}.

\begin{figure*}
  \centering
      \includegraphics[angle=270,width=\textwidth]{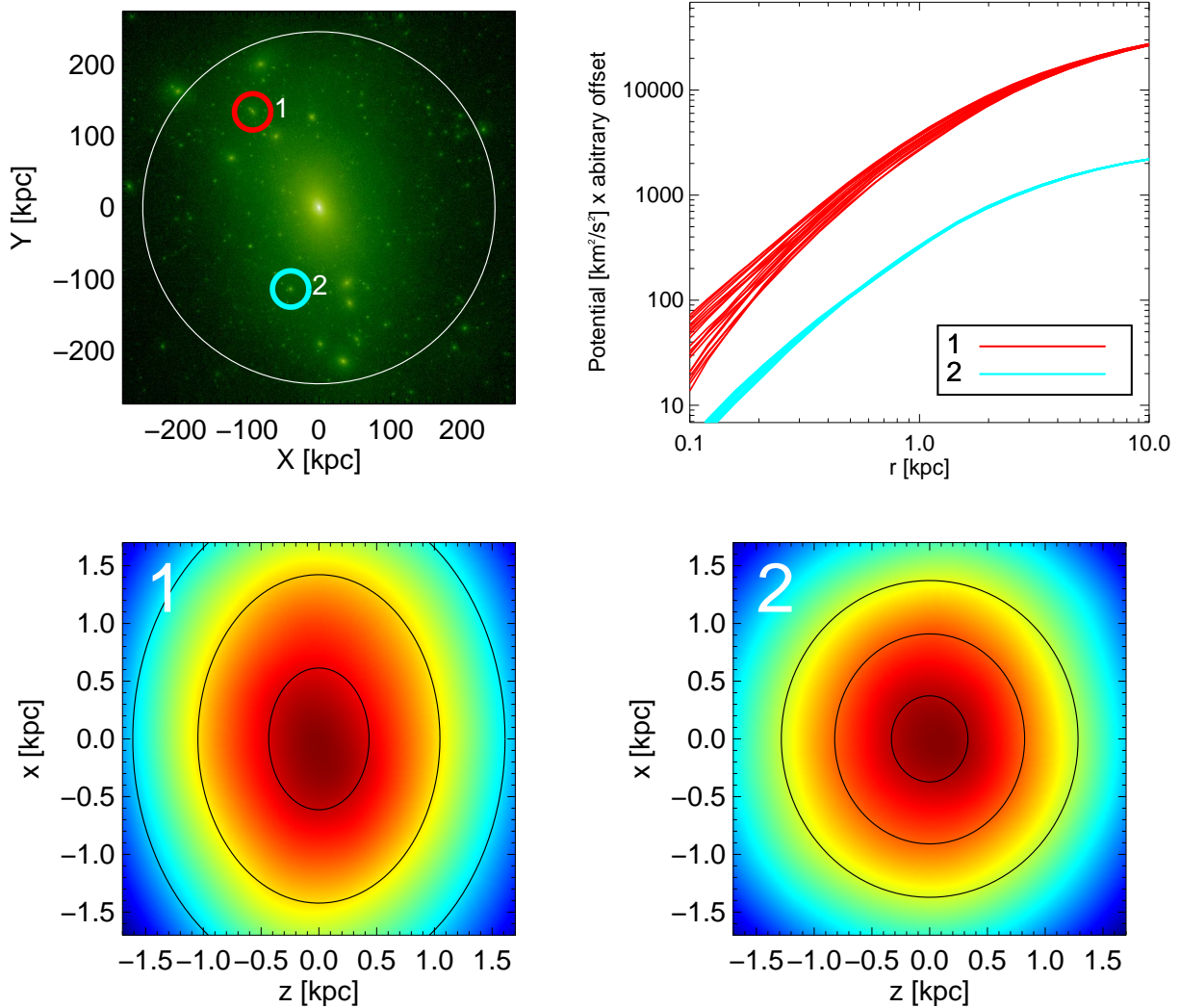}
  \caption{Isopotential contours measured for two subhaloes selected from the Aq-A halo. {\it Top left panel:} Projected density plot of Aq-A-4. The red and cyan circles indicate the positions of the two subhaloes chosen for illustration. {\it Top right panel:} Gravitational potential as a function of radius measured along 100 isotropic directions from the centre of each subhalo. {\it Bottom panels:} Ellipsoidal fits (black ellipses) to isopotential contours superimposed on a 2D slice of their gravitational potentials, in the diagonal frame of each subhalo. The $x$ and $z$ coordinates are aligned with the major and minor axis, respectively. The depth of the potential increases from blue to red.}
  \label{fig:methods}
\end{figure*}

\section{The Numerical Simulations}\label{sec:sims}    

Our analysis uses the Aquarius Project, a suite of $N$-body cosmological simulations of the formation of Milky Way-sized haloes in a $\Lambda$CDM universe \citep{Springel2008}. These state-of-the-art simulations follow the evolution of six dark matter haloes with virial\footnote{We define the virial mass, $M_{200}$, of a halo as that enclosed by a sphere of mean density 200 times the critical density of the Universe, $\rho_{\rm crit} = 3H^2/8\pi G$. Virial quantities are defined at that radius, and identified by a “200” subscript.} masses in the range $0.8$-$1.8 \times 10^{12} \Msun$. The six haloes are named 'Aq-A' through 'Aq-F' (although we exclude Aq-F from this work due to a recent major merger which is not thought to be representative of the formation of the MW), suffixed with various resolution levels ranging from 1 (highest; particle mass $m_p \sim 10^3 \Msun$) to 5 (lowest; $m_p \sim 10^6 \Msun$). These simulations adopt a ``standard'' $\Lambda$CDM cosmogony, with the same parameters as the Millennium Simulation \citep{Springel2005}: $\Omega_{\rm M}=0.25$, $\Omega_{\Lambda}=0.75$, $h=0.73$, $n=1$, and $\sigma_8=0.9$.

Subhaloes of at least 20 particles were identified in each of the main haloes  using {\small SUBFIND} \citep{Springel2001}, a recursive algorithm that locates gravitationally bound overdensities within groups of particles identified by a friends-of-friends (FoF) groupfinder \citep{Davis1985}. For further details we refer the reader to \citet{Springel2008}.

In order to identify subhaloes that are likely to contain dSph galaxies, we use a semi-analytic model grafted onto the $N$-body simulation, as described by \citet{Starkenburg2013}. The model includes prescriptions for star formation, gas cooling, as well as gas heating and ejection via supernova feedback. Also included is the interaction between a satellite and its host, such as stellar and gas stripping due to interactions with the host's tidal field, as well as ram pressure stripping of hot gas upon accretion. 

Unless otherwise specified we use the level-2 version of the Aquarius simulations, which have a particle mass of $\sim 10^4 \msun$ and a Plummer-equivalent force resolution of $\sim 66$ pc. Only systems with at least 700 particles are retained for analysis, in order to ensure numerical convergence (see \se{convergence}). 


\section{Analysis}\label{sec:analysis}

\subsection{Potential-fitting method}\label{sec:method}

Most previous studies have chosen to measure halo shapes by fitting triaxial ellipsoids to the moment of inertia tensor, which is closely related to the density of the system \citep[e.g.,][]{Kuhlen2007, Knebe2010, VeraCiro2014}. One disadvantage of this approach is that isodensity contours are often far from ellipsoidal due to the presence of substructure \citep{Springel2004, Hayashi2007}. 

{The gravitational potential, on the other hand, is much less sensitive to local density variations since its value at a particular point depends on the inner and outer mass profile of the whole halo. As a consequence, isopotential contours are much smoother and better approximated by ellipsoids than isodensity contours.} Furthermore, the dynamics of stars in dSph galaxies are dominated not by the local density, but by the gravitational potential of the dark matter halo. Thus the shape of the potential, rather than the density, should be more relevant to observations and this is therefore the approach we take in this work. The technique we use to measure the gravitational potential of the Aquarius satellite haloes follows that of \citet{Hayashi2007}, where the reader may find details and tests on the method. We only outline briefly the basic procedure here. 

The centre of the halo is first identified as the minimum of the gravitational potential, computed by direct summation over all bound particles given by {\small SUBFIND}\footnote{We have explicitly checked that the exclusion of the host halo only affects the measured shape of the potential beyond the outer 2-3 kpc of typical luminous subhaloes that reside within the virial radius of the main halo, well beyond the small ($ < 1$ kpc) radii pertinent to the main results of this paper.}. We then measure the potential along 100 isotropic radial directions at 20 distances from the centre, logarithmically spaced in the range $0.07$ to $14$ kpc, {to ensure that the potential near the centre of the halo is well-sampled.}

\begin{figure*}
  \centering
      \includegraphics[angle=270,width=\textwidth]{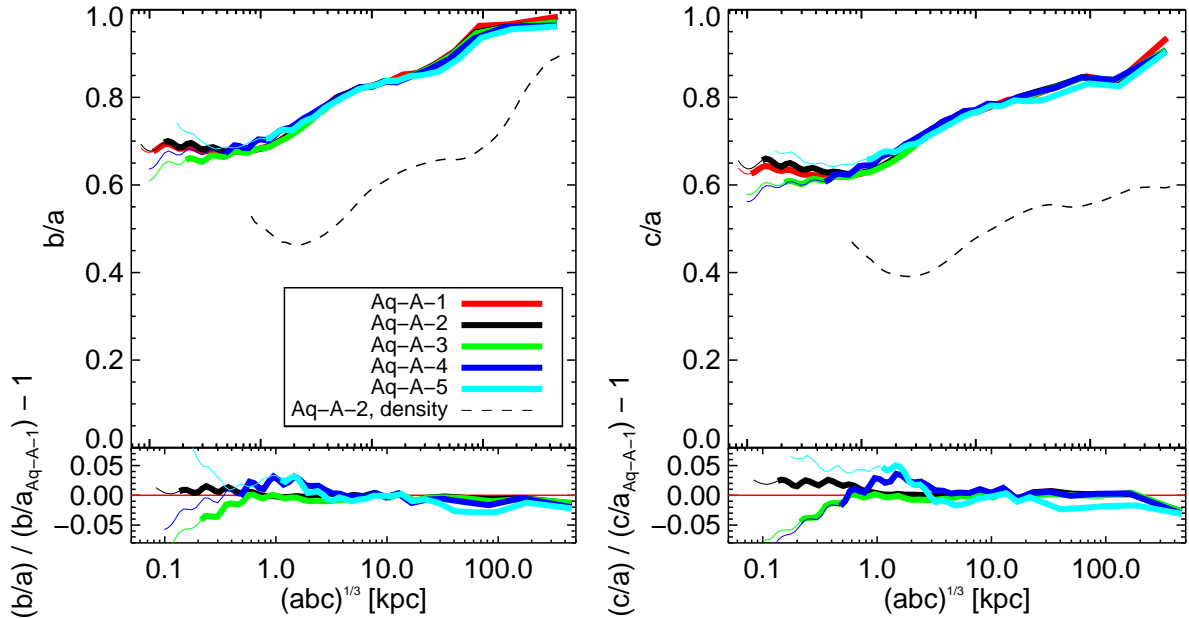}
  \caption{Gravitational potential axis ratios for the Aq-A main halo at five different resolutions. The ratios between intermediate and major axes ($b/a$) and minor and major axes ($c/a$) are shown in the left and right panels, respectively, as a function of the ellipsoidal radius $(abc)^{1/3}$. Residuals relative to the highest resolution run are shown in the bottom panels. Lines become thinner inside our adopted convergence radius, $r_{\rm conv}^{[0.1]}$. Dashed lines indicate the respective axis ratios for isodensity contours in Aq-A-2 taken from \citet{VeraCiro2011}.}
  \label{fig:convergence}
\end{figure*}

\begin{figure*}
  \centering
      \includegraphics[angle=0,width=\textwidth]{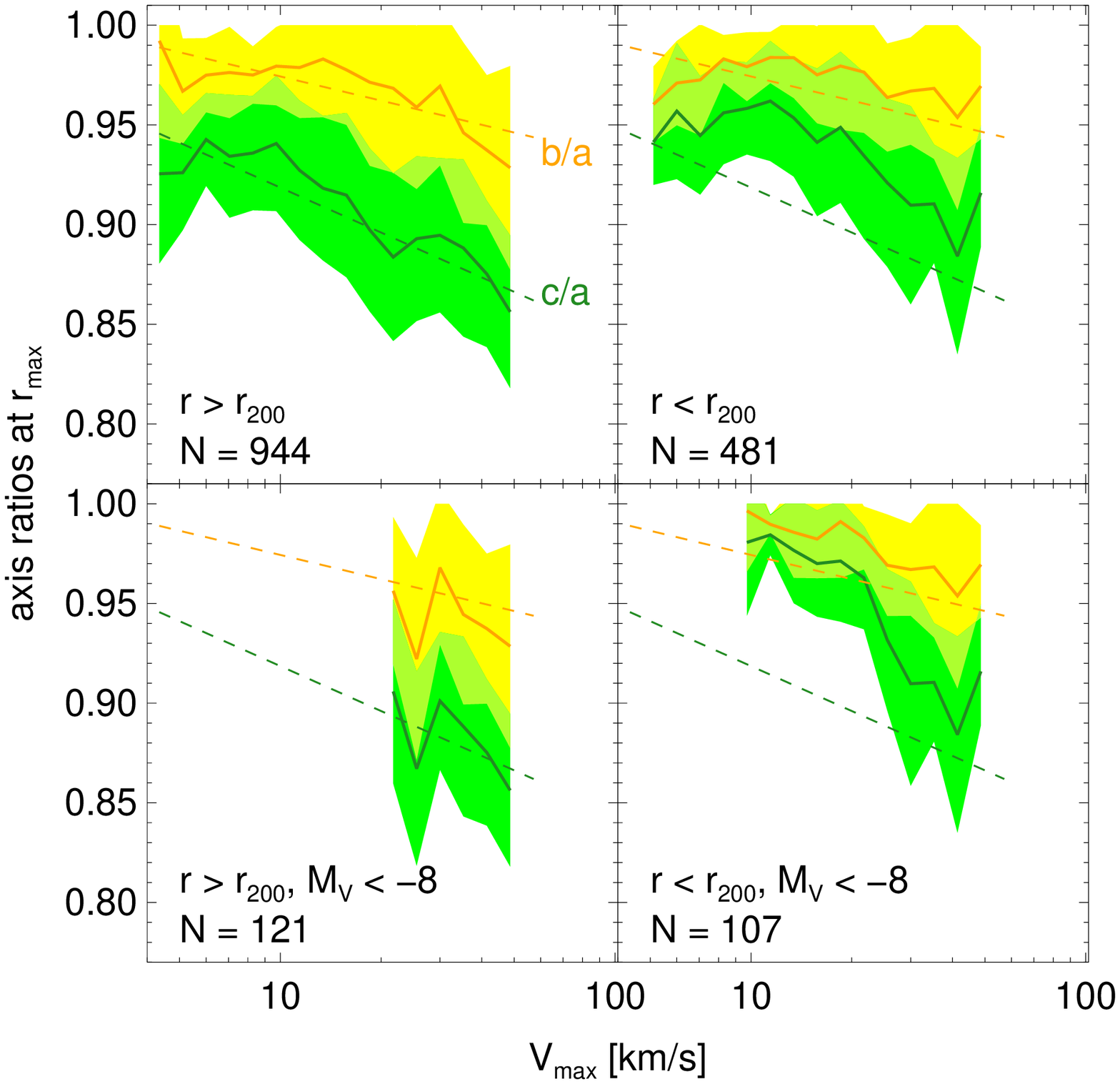}
  \caption{Axis ratios as a function of maximum circular velocity for various haloes in the  mass range corresponding to dwarf galaxies. The left and right columns show haloes in the field (i.e., outside the virial radius, $r >\r200$) and subhaloes of the main halo ($r < \r200$), respectively. The bottom rows correspond to systems with luminous components brighter than $M_V=-8$. $b/a$ and $c/a$ are shown in yellow and green respectively. Solid lines indicate running medians while filled areas indicate the 1$\sigma$ scatter. Overlapping regions are coloured in yellow-green. Fits to the axis ratios in the top left panel are reproduced in all other panels for reference.}
  \label{fig:axisratios_vs_Vmax}
\end{figure*}

{The shape of a given isopotential contour is then found by fitting a 3D ellipsoid to the $100$ points (interpolated along each ray) that match a specified potential value. We find the directions of the major, intermediate, and minor axes by diagonalizing the inertia tensor of these points, assuming that the centre of the ellipsoid lies halfway between their minimum and maximum Cartesian coordinate values. In the diagonal reference frame the ellipsoid axis lengths are computed by minimizing the deviations \be S = \sum_{i=1}^{100} \left( {\hat r}_i - \sqrt{x_i^2 + y_i^2 + z_i^2} \right)^2 \ee from an ellipsoid with major, intermediate, and minor axis lengths $a$, $b$, and $c$, respectively. Here ${\hat r}_i$ is the normalized radius of each isopotential point, defined by \be \frac{x_i^2 + y_i^2 + z_i^2}{{\hat r}_i^2} = \frac{x_i^2}{a^2} + \frac{y_i^2}{b^2} + \frac{z_i^2}{c^2}. \ee As in \citet{Hayashi2007}, we minimize $S$ using a Newton-Raphson method to alternatingly find the roots ${\partial S}/{\partial a}=0$, ${\partial S}/{\partial b}=0$, and ${\partial S}/{\partial c}=0$ until $S$ can not be minimized further. An illustration of the result of our procedure is shown in \Fig{methods} for two subhaloes selected from the Aq-A-4 halo: subhalo 1 is very elliptical; subhalo 2 much less so. Ellipsoids are excellent fits to the isopotential contours in these subhaloes.}

\subsection{Numerical Convergence} \label{sec:convergence}

To investigate the shape of the potential in the central regions of a halo, we need to  understand the robustness of our measured shapes as a function of radius. We have therefore tested the convergence of our algorithm on the Aq-A main halo at different resolution levels. Specifically, we have explored whether the definition of ``convergence radius'', $r_{\rm conv}$, first proposed by \citet{Power2003} holds for isopotential shapes. These authors show that halo mass profiles fail to converge numerically in regions where the local collisional relaxation timescale, $t_{\rm relax}(r)$, is comparable to or shorter than the age of the Universe. Expressing $t_{\rm relax}$ in units of the circular timescale at the virial radius, $t_{\rm circ}(r_{200})$, we have
\be
\begin{split} \kappa(r) & \equiv \frac{t_{\rm relax}(r)}{t_{\rm circ}(\r200)} = \frac{N}{8 \ln N} \frac{r/V_{\rm c}}{\r200/\V200} \\ 
& = \frac{\sqrt{200}}{8} \frac{N(r)}{\ln N(r)} \left( \frac{\overline{\rho}(r)}{\rho_{\rm crit}} \right)^{-1/2} 
\end{split}
\label{eq:rconv}
\ee where $N(r)$ is the number of particles within radius $r$, $\rho_{\rm crit}$ is the critical density of the Universe, and $\overline{\rho}(r)$ is the mean density within $r$. The value of $\kappa$ sets the level of accuracy to which the circular velocity can be estimated outside $r_{\rm conv}^{[\kappa]}$: for example, $\kappa = 1$ identifies the innermost radius, $r_{\rm conv}^{[1]}$, where $V_{\rm circ}(r)$ may be estimated to better than $10\%$. Similarly, outside $r_{\rm conv}^{[7]}$ circular velocities converge to better than $2.5\%$ \citep{Navarro2010}.

We measure the shapes of isopotential contours in the Aq-A main halo at resolutions 1 (highest) through 5 (lowest). The ratios $b/a$ and $c/a$ are shown as a function of the ellipsoidal radius $(abc)^{1/3}$ in \Fig{convergence} (thick lines). For reference, the axis ratios of {\it isodensity} contours of the Aq-A-2 main halo, as measured by \citet{VeraCiro2011},  are plotted with dashed lines down to $r_{\rm conv}^{[7]}$. As expected, isopotential contours are more spherical than the density at all radii.  

\Fig{convergence} also shows excellent convergence in the shape measurements down to radii much smaller than the region where the enclosed mass profile converges. Even at $r_{\rm conv}^{[0.1]}$ (the smallest radii shown in thick line type in \Fig{convergence}) shapes deviate by less than 5 per cent from the highest resolution Aq-A-1 run.  We ascribe this to the fact that the potential is sensitive to the whole halo, rather than just its inner mass distribution. We shall therefore hereafter adopt $r_{\rm conv}^{[0.1]}$ as our fiducial convergence radius when measuring gravitational potential shapes.

\section{dwarf subhalo shapes}\label{sec:shapes}

\subsection{Shape vs mass}

We begin by studying the dependence of the gravitational potential shape on halo mass, in the regime appropriate for dwarf galaxies; $10^7 \Msun \simlt M_{200} \simlt 5 \times 10^9 \Msun$, or $4$ km/s $<\Vmax <35$ km/s \citep[see, e.g.,][]{Starkenburg2013,Barber2014}. Here $\Vmax$ is the maximum halo circular velocity (reached at radius $r_{\rm max}$), a robust measure of halo mass that is less sensitive to the definition of the outer boundary of a subhalo and to its location within the parent halo \citep[see, e.g.,][]{Penarrubia2008}.

 \Fig{axisratios_vs_Vmax} shows the axis ratios, measured at $r_{\rm max}$, of all dwarf galaxy haloes in the level-2 Aquarius runs, split between subhaloes (i.e., haloes identified by {\small SUBFIND} within the virial radius of the main halo) and isolated ``field'' haloes (i.e., the central halo of each separate FoF group beyond $r_{200}$). The solid curves track the median of the distribution, whereas the dashed lines are simple fits to the full field halo sample, reproduced in each panel for ease of comparison. 

The top-left panel of \Fig{axisratios_vs_Vmax} shows that field haloes become less spherical with increasing mass, confirming prior results reported in the literature \citep[see, e.g.,][and references therein]{VeraCiro2014}. At a given $\Vmax$, subhaloes are generally more spherical than field haloes, a result that is especially clear for the $c/a$ ratio (compare the green bands in the top left and right panels of \Fig{axisratios_vs_Vmax}).

The bottom panels of the same figure show results for a subsample of haloes deemed, according to the semi-analytical model, to host relatively luminous satellites (i.e.,  ``classical'' dSphs, defined by the condition $M_V < -8$). We shall hereafter refer to these as ``luminous'' or ``classical'' haloes or subhaloes, respectively. As shown in the bottom-left panel, luminous field haloes populate preferentially the high-mass end of the distribution: no field halo below $\Vmax \sim 20$ km/s hosts a satellite brighter than $M_V=-8$. 

On the other hand, subhaloes that host classical dwarfs differ systematically from their field counterparts in both $\Vmax$ and shape, as shown in the bottom-right panel of \Fig{axisratios_vs_Vmax}. Even low mass subhaloes (under $20$ km/s) host some of those satellites, a result of tidal stripping. In addition, like all subhaloes, they are substantially rounder than their field counterparts, especially at the low $\Vmax$ end where tidal stripping has been the strongest. This suggests a direct link between tides and the sphericalization of subhaloes, an effect that has been hinted at in earlier work \citep[e.g.][]{Kuhlen2007}. We explore it further below after checking first that our results are not unduly sensitive to the radius chosen to measure shapes.

\begin{figure}
  \centering
  \includegraphics[angle=0,width=0.5\textwidth]{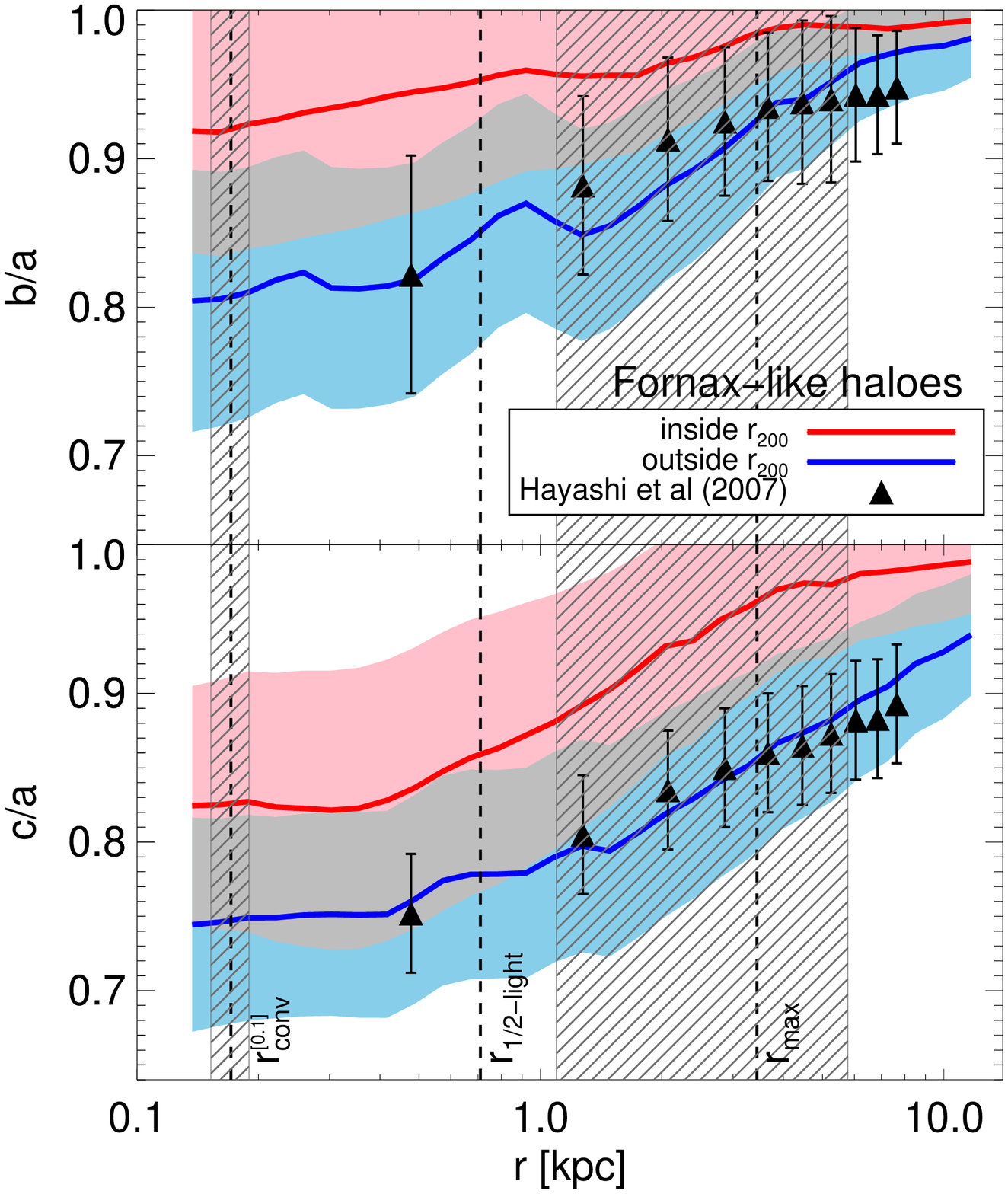} 
\caption{Axis ratios as a function of radius for Fornax-like haloes. Isolated haloes (i.e., outside $r_{200}$) are shown in blue, subhaloes in red. Solid lines indicate medians, filled areas the rms scatter. Overlapping regions are plotted in grey. The black triangles indicate the results of \citet{Hayashi2007} for isolated halo resimulations, for reference. Shaded vertical bands show the convergence radius and $\rmax$ of Fornax-like haloes. The half-light radius of Fornax itself is indicated with a vertical dashed line.}
  \label{fig:ba_vs_r_Fornax}
\end{figure}

\subsection{Radial dependence}

Although $r_{\rm max}$ is a well-defined characteristic radius of the dark matter profile, we would like to apply our shape measurements to the innermost regions, typically well within $r_{\rm max}$, inhabited by the stellar component. We therefore explore the radial dependence of our results in \Fig{ba_vs_r_Fornax} for the case of luminous systems whose predicted model luminosities are similar (within $0.5$ mag) to Fornax.

The red and blue lines indicate the median axis ratios as a function of radius for this Fornax-like subpopulation split between subhaloes and isolated haloes, respectively. For isolated haloes (shown in blue) our results are consistent with earlier work. Haloes are generally triaxial near the centre, with $b/a \sim 0.8$ and $c/a \sim 0.75$, and become gradually more spherical outwards, as expected from the increasing monopole dominance in the outer regions. The solid triangles indicate the shape-radius relation reported by \citet{Hayashi2007} for isolated haloes, scaled to our mean value of $r_{\rm max}$. The agreement between their results and ours is remarkably good.

Note as well that the red curves lie above the blue curves at all radii, indicating that the conclusions of the previous subsection are robust: subhaloes are more spherical than isolated haloes regardless of where shapes are measured. Finally, the same figure shows that our measurements can be applied to radii comparable to the size of the stellar component of dSphs: our fiducial convergence radius, $r_{\rm conv}^{[0.1]}\sim 160$ pc, is well within the $\sim 700$ pc half-light radius of dwarfs like Fornax.

\subsection{Shape vs tidal stripping}

The results of the previous subsections strongly suggest that haloes become more spherical as they lose mass to tidal stripping. This is somewhat counterintuitive, since elongated shapes are one of the signature features of tidal interactions. Those features, however, occur during a short-lived phase just after a galaxy has gone past the pericentre of its orbit, where the tides are strongest. Once the tails of escaping particles have moved off, the bound remnant quickly relaxes to equilibrium and settles into a less triaxial configuration. At apocentre, where a galaxy spends most of its evolution, little evidence of the tidal interaction typically remains in the vicinity of the bound remnant \citep[see, e.g.,][for details]{Penarrubia2008}.

\begin{figure}
  \centering
      \includegraphics[angle=0,width=0.5\textwidth]{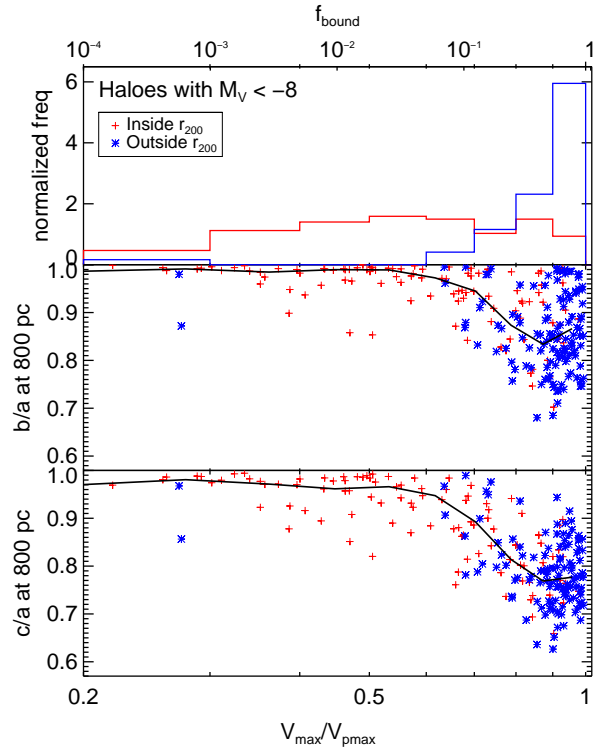}
  \caption{The ratio, $V_{\rm max}/V_{\rm pmax}$, between maximum circular velocity at $z=0$ and its peak value in the past is used as a measure of the importance of tidal stripping (x-axis). The scale on top shows the corresponding mass fraction retained by the bound remnant. {\it Top panel:} Distribution of $V_{\rm max}/V_{\rm pmax}$ for all 228 ``classical'' (i.e., $M_V<-8$) haloes in our sample. Systems inside and outside the virial radius are shown in red and blue, respectively. {\it Middle and bottom panels:}  Axis ratios measured at 800 pc as a function of stripping.  Black lines indicate running medians.}
  \label{fig:axisratios_vs_stripping}
\end{figure}

We show the effect of tidal stripping on shapes more explicitly in the bottom two panels of \Fig{axisratios_vs_stripping}. Here we plot the axis ratios, measured at a fiducial radius of $800$ pc, for all haloes likely to host ``classical'' dwarfs, as a function of the degree to which they have been stripped. The latter is estimated from $\Vmax / \Vmaxmax$, the ratio between its current maximum circular velocity and the peak value it achieved in the past.

This ratio is a sensitive measure of tidal mass loss for $\Lambda$CDM haloes, which are well approximated by the Navarro-Frenk-White profile \citep[hereafter NFW,][]{Navarro1996, Navarro1997}. Only haloes severely affected by tides see a noticeable reduction in $\Vmax$: a halo that has lost $90\%$ of its original mass sees its $\Vmax$ reduced, on average, to $\sim 70\%$ of its original value. A rough conversion between the remaining bound mass fraction ($f_{\rm bound}$) and $\Vmax / \Vmaxmax$ is provided by the scale on the top axis of \Fig{axisratios_vs_stripping} \citep[see][for more details]{Penarrubia2008}.

The top panel of \Fig{axisratios_vs_stripping} shows clearly that the ``classical'' subhalo population (in red) is more heavily stripped than their counterpart haloes found at present outside\footnote{Note that being outside the virial radius does not necessarily imply that a halo has been unaffected by tides. Indeed, many haloes in the outskirts, e.g., $1<r/r_{200}<2$, are on highly-energetic orbits and have already been inside the halo virial boundary \citep[see, e.g.,][for details and further references]{Barber2014} } the virial radius (in blue). Heavy stripping leads to strong sphericalization; indeed, haloes that have lost more than $90\%$ of their mass are essentially spherical, with median $b/a$ and $c/a$ exceeding $\sim 0.95$. On the other hand, subhaloes that have not been stripped substantially are as triaxial as their isolated counterparts. This confirms that being a subhalo is not sufficient to ensure sphericalization; heavy tidal stripping seems needed in order to prompt a significant change in shape.

\begin{figure*}
  \centering
      \includegraphics[angle=0,width=\textwidth]{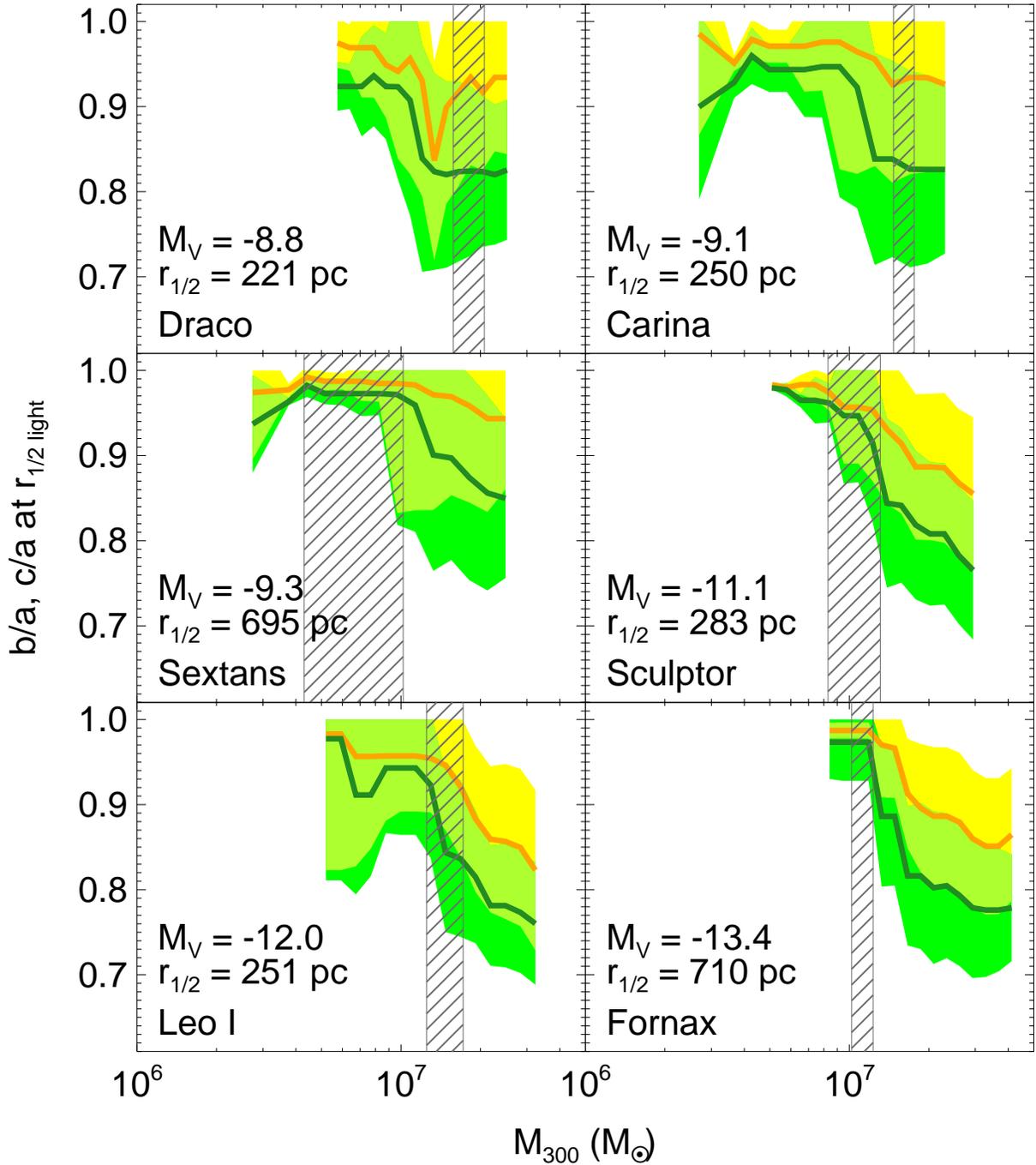}
\caption{ Axis ratios of subhaloes with luminosities matching ($\pm 0.5$ mag in $V$) that of six Milky Way dSphs, shown as a function of the total mass enclosed within 300 pc. Axis ratios are measured at the half-light radius of each MW dwarf, as indicated in the legends. $b/a$ and $c/a$ are shown in orange and green respectively. Solid lines indicate running medians while filled areas indicate the $1 \sigma$ scatter. $M_{300}$ constraints shown by the dashed vertical bands are taken from \citet{Strigari2008}. }
  \label{fig:axisratios_vs_M300_sats}
\end{figure*}

\begin{figure}
  \centering
      \includegraphics[angle=270,width=0.5\textwidth]{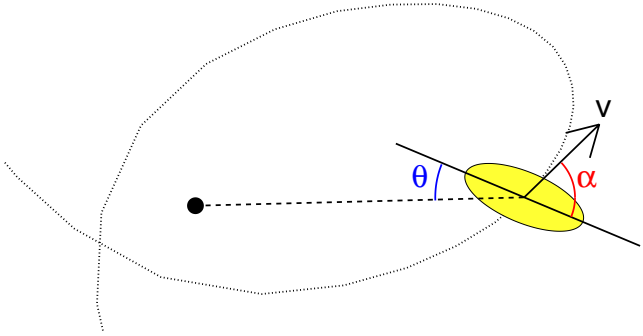}
  \caption{Schematic illustration of the angle definitions used in \se{orientation}. The black filled circle indicates the centre of the main halo, while the instantaneous direction of the subhalo velocity is shown as an arrow. The subhalo trajectory in the orbital plane is shown by the dotted curve. The figure assumes, for simplicity, that the major axis of the subhalo lies on the plane of the orbit. }
  \label{fig:toymodel}
\end{figure}

\begin{figure}
  \centering
      \includegraphics[angle=0,width=0.5\textwidth]{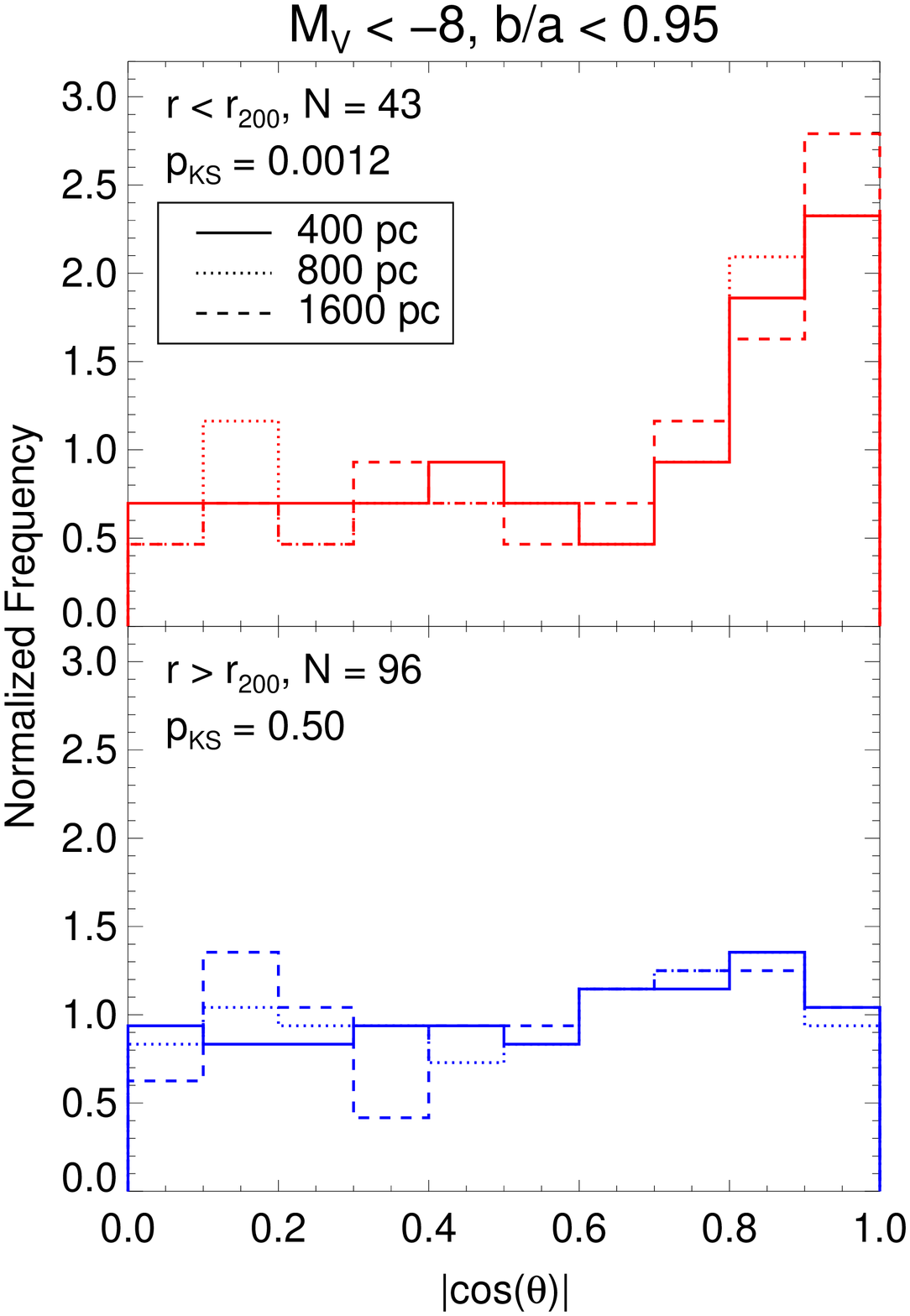}
  \caption{ Distribution of the cosine of the angle between the major axis of a halo/subhalo and the direction to the centre of the main halo, for all ``classical'' ($M_V < -8$) haloes. Subhaloes ($r < \r200$) and field haloes ($r > \r200$) are shown in the top and bottom panels, respectively. The major axis direction is measured at three different radii from the centre of each system: $400$, $800$, and $1600$ pc. The corresponding distributions are shown as solid, dotted, and dashed histograms, respectively. Note the clear radial alignment of the subhaloes. Field haloes, on the other hand, are consistent with a uniform distribution (random orientation). The KS probability of the distribution (at $800$ pc) compared with random orientations is listed in the legends.  }
  \label{fig:orientation_relmain_600}
\end{figure}

\subsection{Implications for MW satellites}

The results mentioned above have interesting implications for the Milky Way dSph satellites, where, except for Sagittarius, it has been difficult to reach unambiguous conclusions regarding the importance of Galactic tides on their structure and evolution. A heavily stripped satellite should be nearly spherical, but it would also be less massive, on average, than a system of similar luminosity that has evolved unaffected by tides---mass stripping should affect first and foremost the outer, mainly dark component. In other words, the potential of systems that are more luminous than expected given their total dark mass should be nearly spherical if their dark matter content has been depleted by tides.

We apply this idea to six MW satellites in \Fig{axisratios_vs_M300_sats}, where we plot $M_{300}$, the mass enclosed within $300$ pc, versus axis ratios for all haloes and subhaloes with luminosities comparable to Draco, Carina, Sextans, Sculptor, Leo I and Fornax, respectively, at $z=0$ as determined by our semi-analytic model. Estimates of $M_{300}$ are taken for all of these dwarfs from \citet{Strigari2008}, and are shown by the shaded areas in \Fig{axisratios_vs_M300_sats} \citep[Note that these values assume spherical symmetry in the dark matter haloes whereas we predict haloes to be triaxial in nature; however they should be sufficiently accurate for this qualitative analysis; see also][]{Hayashi&Chiba2012}. The $M_{300}$ dependence of the potential shapes is very similar to that shown in \Fig{axisratios_vs_Vmax}. This is expected if, to first order, satellites of given luminosity formed in haloes of similar mass (and therefore similar $M_{300}$) that have subsequently been stripped to various degrees.

Take, for example, Fornax, one of the most luminous of the classical dSphs but also one where, given its relatively low mass-to-light ratio, dark matter is less prevalent than in others. Because of its luminosity, the semi-analytic model assigns most Fornax analogues to relatively massive haloes\footnote{Recall that the semi-analytical model takes into account stellar stripping as well as the stripping of the dark matter halo, so some Fornax analogues could have been much brighter than Fornax at earlier times}. Such haloes, if unaffected by tides, would have masses of order $4 \times 10^7 M_\odot$ within $300$ pc, greatly exceeding the estimate of \citet{Strigari2008} (see bottom-right panel of \Fig{axisratios_vs_M300_sats}). Only heavily-stripped (and therefore essentially spherical) subhaloes would be consistent with this semi-analytic model of Fornax, {a result that may help to resolve the ``too-big-to-fail problem'' of \citet{BoylanKolchin2011}}. Constraints on the shape of the potential of Fornax would therefore be invaluable to assess the validity of our semi-analytic approach.

At the other extreme, Draco is a faint dSph with a large mass-to-light ratio that is easily accommodated in the model if tidal stripping has been, at most, modest (see top-left panel of \Fig{axisratios_vs_M300_sats}). In that case, our model predicts that Draco's potential should be significantly triaxial. The same procedure suggests that Sextans and Sculptor should inhabit heavily stripped---and thus nearly spherical---subhaloes, whereas Carina and Leo I should be relatively unaffected by tides and hence triaxial. We plan to explore the consistency of this prediction with a wider array of independent constraints in a future contribution. 

\section{Orientation of dwarf galaxy satellites}\label{sec:orientation}

Although tidal stripping renders subhaloes more spherical, it is important to note that many of them have been only modestly affected, so that the orientation of their principal axes is still well defined. As discussed in Sec.~\ref{SecIntro}, earlier work has reported a significant alignment between the major axis of a subhalo and the radial direction to the main halo. We extend this work here by focusing on the innermost regions of the halo, where the luminous component of the galaxy is expected to reside. Our simulations also allow us to shed some light into the origin of such alignment and to make predictions for the satellite population of a galaxy like M31 that can be contrasted directly with observations.

\begin{figure*}
  \centering
      \includegraphics[angle=270,width=\textwidth]{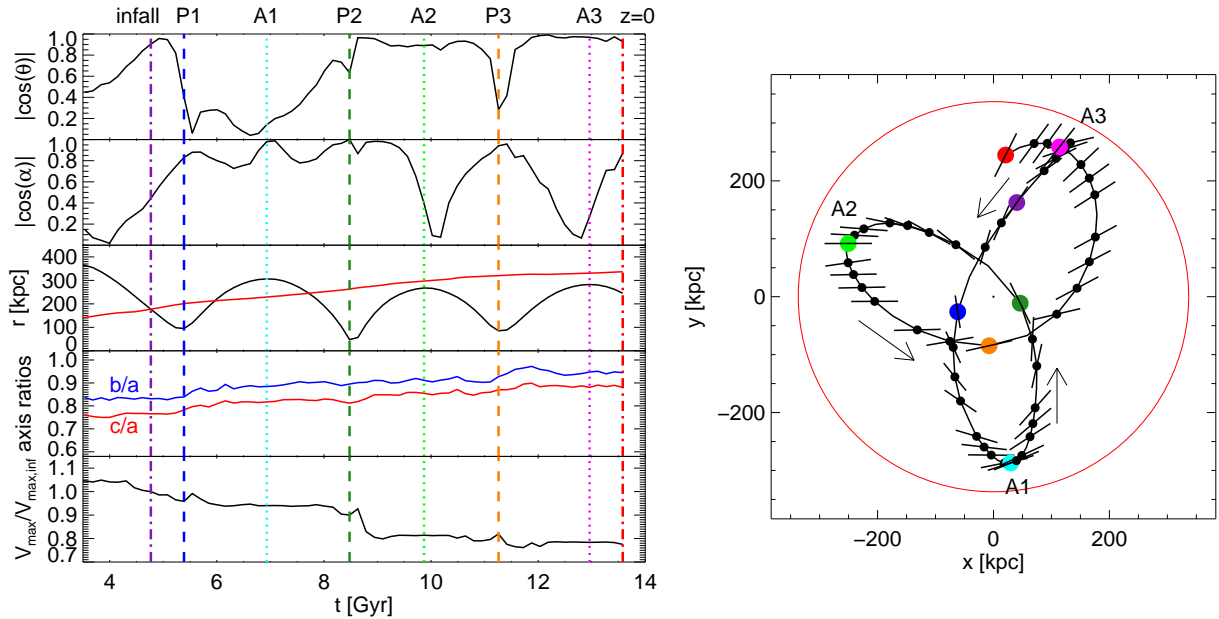}
  \caption{ Evolution of a subhalo that is radially aligned at $z=0$. The left panels show, as a function of time, the evolution of several parameters that measure, from top to bottom: radial alignment; orbital stretching; distance to main halo centre; axis ratios at 800 pc; and degree of stripping since infall, respectively. The solid red line in the middle panel is the evolving virial radius of the main halo. Vertical lines indicate characteristic times in the orbit; pericentres and apocentres are dashed and dotted, respectively.  The right panel shows the orbit of this subhalo from $t=4$ Gyr until $z=0$, projected onto the orbital plane at $z=0$. Black dots and lines indicate the direction of the subhalo's major axis projected onto the plane of the orbit. Each coloured dot corresponds to the time indicated by the vertical line of the same colour in the left panels. Arrows indicate the subhalo's direction of motion. The red circle indicates the virial radius of the main halo at $z=0$.}
  \label{fig:orbitexample_aligned}
\end{figure*}

\subsection{Subhalo radial alignment}\label{sec:align_main}

If subhaloes were oriented randomly, the distribution of the cosine of the angle, $\theta$, between its major axis and the direction to the centre of the main halo (see \Fig{toymodel}), should be uniform.  This is indeed the case for ``isolated'' haloes identified outside the virial radius, but not so for subhaloes, as shown by the bottom and top panels of \Fig{orientation_relmain_600}, respectively. 

We include in this figure all systems deemed to host a ``classical'' dSph and whose intermediate-to-major axis ratio does not exceed $0.95$. We have carefully tested on artificially flattened systems that only in such systems is the major axis direction accurately determined. We have also explicitly checked that orientations are insensitive to the precise subset of particles used to compute the potential: re-measuring potentials using only particles within $0.5$, $1$, or $2\, r_{\rm max}$ yields indistinguishable results to those obtained using all particles identified by {\small SUBFIND}.

The top panel of \Fig{orientation_relmain_600} shows the distribution of $|\cos(\theta)|$ for major axis orientations measured at $400$, $800$, and $1600$ pc from the centre of ``classical'' subhaloes ($43$ in total for all Aquarius runs). The distribution is clearly non-uniform and shows a strong preference for the major axis to align with the radial direction, in agreement with earlier work. A KS test reveals this distribution is significantly different from uniform with a $p$-value ($p_{\rm KS}$) of 0.0012. The results are insensitive to where the potential is measured, implying that that the principal axis directions are internally well aligned. On the other hand, there is no discernible alignment for ``isolated'' haloes (bottom panel of \Fig{orientation_relmain_600}, $p_{\rm KS} = 0.50$). We conclude that the radial alignment of subhaloes is a result of the tidal forces exerted on them by their host galaxy.

\subsection{Alignment origin}

\begin{figure*}
  \centering
      \includegraphics[angle=0,width=\textwidth]{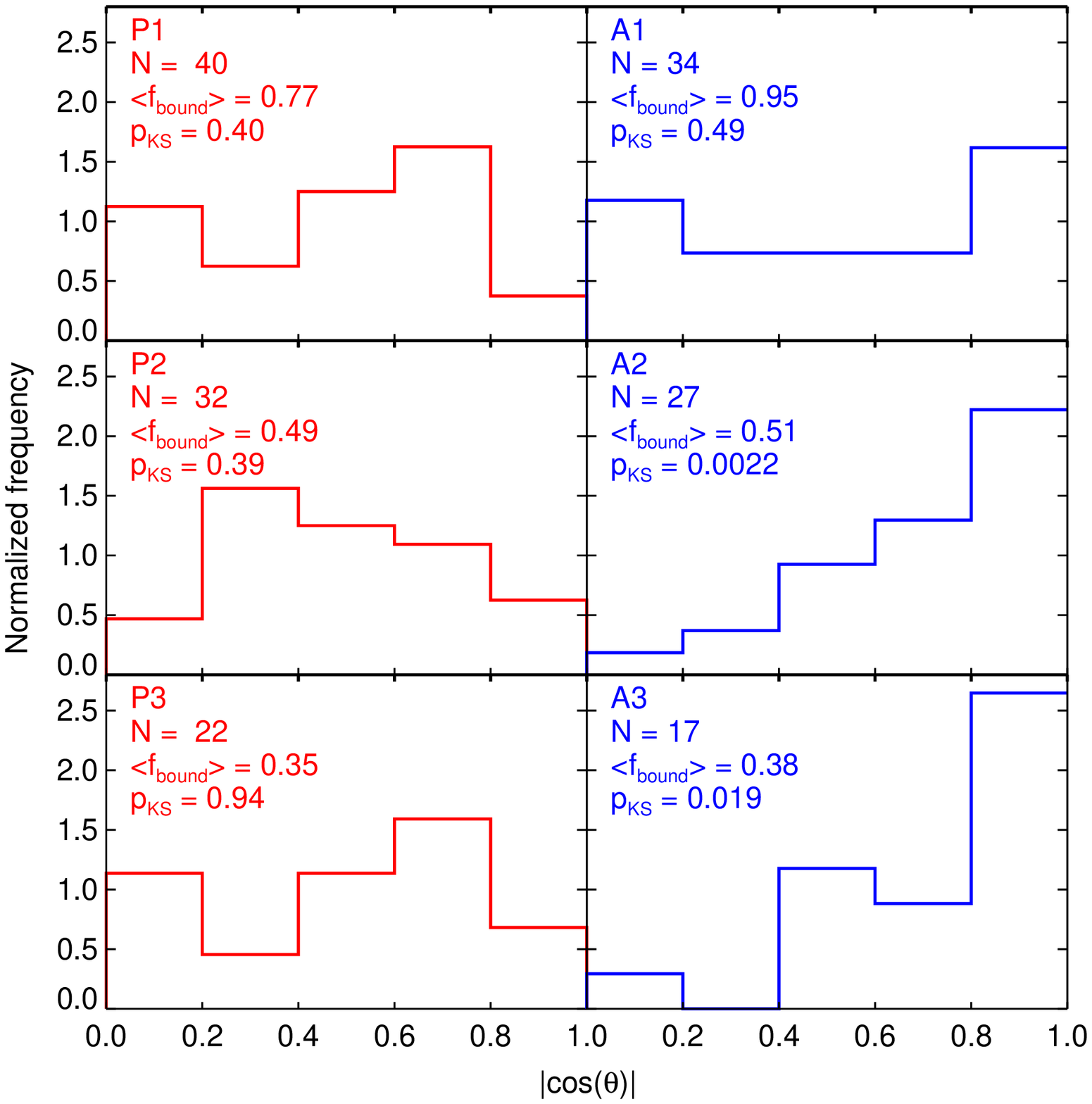}
  \caption{  Distribution of the cosine of the angle between the major axis measured at 800 pc and the radial direction for subhaloes included in \Fig{orientation_relmain_600}, at various orbital phases. P$n$ and A$n$ correspond to pericentres and apocentres, respectively, with the number $n$ indicating 1st, 2nd, etc,  as in \Fig{orbitexample_aligned}. The number of subhaloes, the average fraction of mass that remains bound to each subhalo, and the probability that the sample is drawn from a uniform distribution are indicated in each panel.}
  \label{fig:orientation_hist_apo_peri}
\end{figure*}

We investigate the origin of the alignment seen in the top panel of \Fig{orientation_relmain_600} by tracking each of the $43$ subhaloes back in time in order to pinpoint when and where a subhalo becomes aligned.  

If tidal deformation/stripping is the main cause of the alignment, then subhaloes would be stretched mainly along their orbits, an effect that may produce a radial alignment for very eccentric orbits. Such orbital stretching should yield a detectable preference for the instantaneous major axis direction to align with the orbital velocity. We explore this by monitoring the evolution of the angle $\alpha$ between those two directions (see \Fig{toymodel}) as a function of time for each of the $43$ classical subhaloes. {A value $|\cos(\alpha)| = 1$ throughout the orbit would mean that the longest axis of the subhalo is along its orbit, and would be the signature of this tidal deformation mechanism.}

Alternatively, the alignment might be the result of tidal torquing, which, regardless of orientation, tries at all times to ``twist'' the principal axis of a subhalo and align it with the (radial) direction of the tidal force. In this ``tidal locking'' scenario one would expect the radial alignment to persist during the orbit {(i.e. $|\cos(\theta) = 1|$)}, regardless of its phase.

We explore these two alternatives in the case of a classical subhalo that, by $z=0$, has been able to complete three full orbits and has therefore been substantially affected by tides. Because of its long exposure to the tidal field and its substantial ($\sim 80\%$) mass loss this subhalo represents an ideal test of the two scenarios proposed above. We show this example in \Fig{orbitexample_aligned}. 

The left panel of this figure shows, from top to bottom, respectively, $|\cos(\theta)|$; $|\cos(\alpha)|$; the distance from the main halo centre; the axis ratios measured at $800$ pc; and the degree of tidal stripping. Various characteristic moments in the satellite's orbit are indicated by vertical lines. Apocentric and pericentric passages are defined as A$n$ and P$n$ respectively, where $n = 1,2,3$ correspond to the first, second, and third passages, respectively. In the right panel we plot the orbit of the satellite projected onto its orbital plane at $z=0$. Pericentric and apocentric passages are marked with coloured dots matching the colours of the vertical lines of the left panel.

Two different stages may be identified in the subhalo alignment process. During its first orbit (from P1 to P2) the subhalo is not aligned radially, but rather with the orbital direction (i.e., $|\cos(\alpha)|\approx 1$). This is consistent with the tidal stretching scenario discussed above, and seems to hold during first approach, before any substantial mass loss due to tides occurs.

The situation, however, changes abruptly after the second pericentre (P2), where the subhalo undergoes sudden and substantial mass loss. After P2 the halo remains radially aligned during most of its orbit, as expected in the tidal locking scenario described above, except during its next pericentric passage, when it aligns briefly with the tangential direction. This is suggestive of a resonance between the radial orbital period and the figure rotation of the subhalo, which lends further support to the tidal locking scenario.

\begin{figure*}
  \centering
      \includegraphics[angle=270,width=\textwidth]{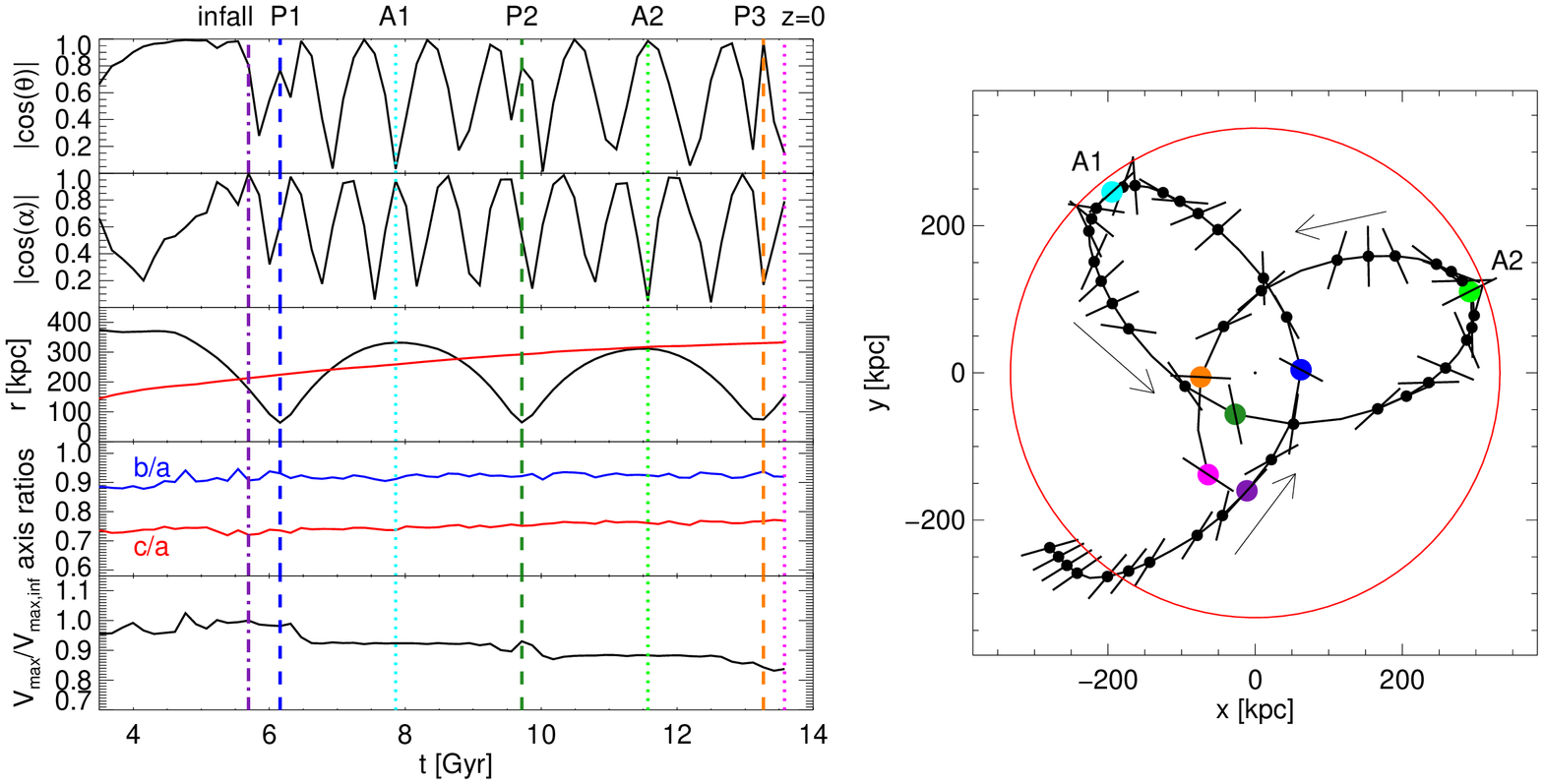}
  \caption{As \Fig{orbitexample_aligned} but for a subhalo whose figure rotation is in the opposite sense to its orbit and that does not align with the radial direction. }
  \label{fig:orbitexample_tumbler}
\end{figure*}

\begin{figure*}
  \centering
      \includegraphics[angle=270,width=\textwidth]{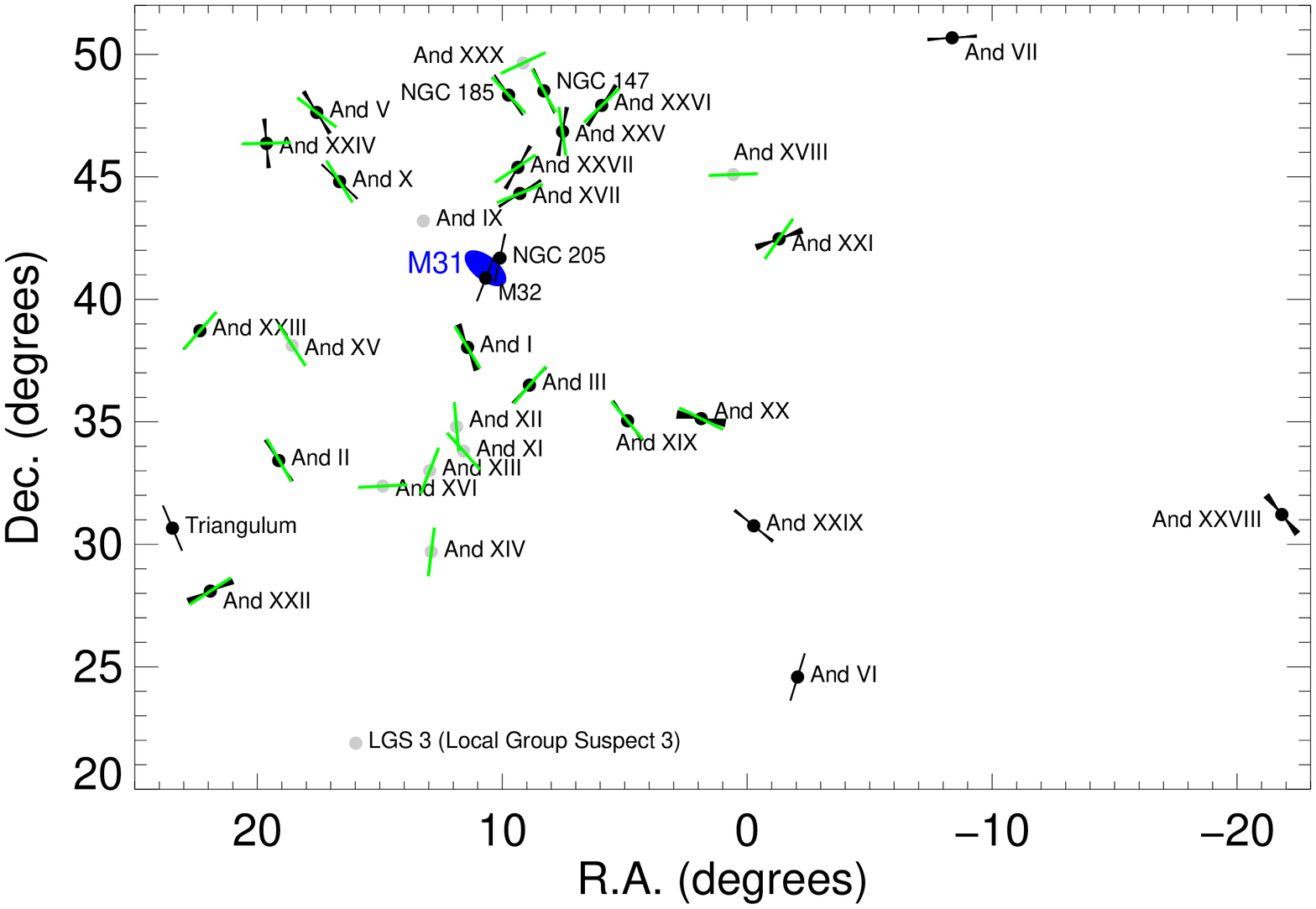}
  \caption{Position and major angle orientation of M31 and its known satellite galaxies taken from the measurements of Martin et al. (2015, in preparation), in green, and the compilation of \citet{McConnachie2012}, in black, from data taken from \citet{deVaucouleurs1991, Choi2002, McConnachie&Irwin2006, Ibata2007, McConnachie2008, Martin2009, Brasseur2011b, Richardson2011, Bell2011, Slater2011}. Errors in position angle are shown as bow ties. Errors from Martin et al. (2015, in preparation) are not shown for clarity. Satellites with currently unpublished or undefined position angles are shown in grey.}
  \label{fig:m31sats_positions}
\end{figure*}

In this interpretation, the subhalo figure rotation is largely tidally-induced and adjusts to the orbital period so as to minimize the torque, which essentially vanishes when $\theta$ is either $0$ or $90$ degrees. By adjusting its figure rotation so as to be radially aligned at apocentre and tangentially aligned at pericentre, the subhalo settles into an orbit where the torque is minimized at both pericenter and apocenter. Note that the torque would also be minimized if the subhalo remained radially aligned at all times, but this would require it to rotate twice as fast; the torque is apparently not strong enough for that, leading the subhalo to settle into a lower order resonance.

This particular example suggests that both the tidal stretching and tidal locking scenarios might hold to some extent, but perhaps during different phases of a subhalo's orbit.  We can test this further by measuring the subhalo alignment not at a given time, but at fixed orbital phase; for example, at different pericentric and apocentric passages. We show this in \Fig{orientation_hist_apo_peri}, where we plot the distribution of $|\cos(\theta)|$ at the first three pericentres (left column) and apocentres (right column). (Note that at apocentre/pericentre $\alpha$ and $\theta$ are complementary, so the distribution of $|\cos(\theta)|$ fully characterizes that of $|\cos(\alpha)|$ and vice-versa.) 

If tidal stretching were important then we would expect the distributions to peak at $|\cos(\theta)|=0$ at apocentric and pericentric passages but there is little evidence for that in \Fig{orientation_hist_apo_peri}. On the other hand, in the tidal locking scenario the alignment should first become evident at A2 and persist at later apocentric passages, whilst vanishing at P2 and subsequent pericentres. The alignment is clearly evident and significant only at A2 and at A3, but there is little evidence for significant departures from random orientation at other times (the legends in each panel of \Fig{orientation_hist_apo_peri} express the KS statistic comparing each distribution to a uniform one.) The data in \Fig{orientation_hist_apo_peri} seem to strongly favour tidal locking as the origin of the radial alignment. {Indeed, we have investigated in detail the 27 subhaloes that complete at least two apocentric passages, finding that 14 of them are clearly tidally locked (meaning that the average $|\cos(\theta)| > 0.8$ over at least one full orbit), except for short dips at pericentres.}

Our results are therefore consistent with the idea that tidal torques operating on a subhalo act to ``spin up'' (or down) its figure rotation so as to lock it into a resonance with its orbital period. The process takes some time, however, and alignments become apparent only for systems that have completed at least one full orbit. This interpretation also implies that tidal locking should operate more effectively in systems with slow figure rotation, which would be easier to spin up. 

Indeed, subhaloes whose patterns rotate too rapidly, or perhaps even counter-rotate relative to their orbits, would be much harder to ``lock'' and should therefore be relatively unaffected. \Fig{orbitexample_tumbler} provides one such example.  This subhalo is tumbling rapidly and in the {\it opposite} direction to its orbit due to a close encounter with another subhalo before P1. As a result, the tide is unable to operate effectively and the subhalo shows no obvious alignment, neither in $\theta$ nor in $\alpha$, during its subsequent orbit. Animations of the evolution of this and other subhaloes may be found at \url{https://sites.google.com/site/barber2014movies/ }.

\subsection{Application to M31 and its satellites}\label{sec:andromeda}

The discussion of the previous section has potentially interesting observational consequences. It implies, for example, that a significant alignment between the position angle of satellites and the radial direction to their host might be detectable observationally. If such an alignment is detected, then it would be a clear indication that the satellite population has completed a number of orbits and that it has been substantially affected by tides. {The discussion below assumes that the stellar component of dwarf spheroidals shares the same orientation as that of their surrounding subhaloes \citep[see, e.g.,][for a recent discussion]{vanUitert2012}.}

In our own Galaxy, our vantage point precludes detection of a radial alignment for the MW dSphs, since measuring their radial extent along our line-of-sight (which almost coincides with the radial direction from the Galactic Centre) is extremely difficult. Thus, alternatively, we apply our results to the satellite population of our nearest large neighbour, the Andromeda galaxy. In \Fig{m31sats_positions} we show the orientation of each satellite relative to M31, using short black lines to indicate the projected position angles from the compilation of \citet[][see \Fig{m31sats_positions} for references]{McConnachie2012}.

\begin{figure}
  \centering
      \includegraphics[angle=270,width=0.5\textwidth]{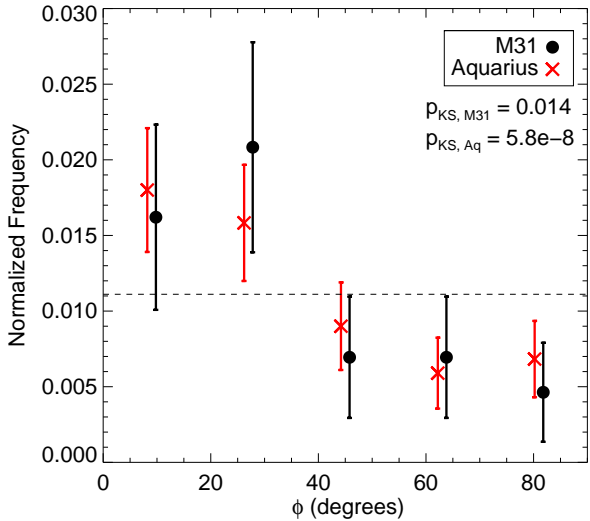}
  \caption{Distribution of the angle between the {\it projected} major axis and radial direction for Aquarius subhaloes brighter than $M_V=-6$ (red crosses) and for M31 satellites (black circles). Note that the radial alignment is also noticeable in projection as an excess of systems with low values of $\phi$ relative to uniform. Both Aquarius subhaloes and M31 satellites seem to show similar alignment. Error bars indicate Poisson noise. The KS probability of each sample compared to uniform is listed in the legend.}
  \label{fig:phi_hist}
\end{figure}

In order to compare with our simulations we compute the alignment expected {\it in projection} for a population of satellites that, according to our semi-analytic model, are brighter than $M_V=-6$, the faintest M31 dwarf for which relevant data are available. Our procedure selects for each subhalo all particles within an ellipsoidal volume equivalent to that of a sphere of radius $800$ pc and uses their projected positions to define principal axes in projection by diagonalizing the corresponding 2D inertia tensor. The result of this exercise (combining three independent projections of each Aquarius halo) is shown in \Fig{phi_hist}, where the red crosses indicate the distribution of $\phi$, the (acute) angle between the projected major axis of a subhalo and the radial direction. The radial alignment discussed in the previous subsections is clearly seen here as an excess of galaxies with low values of $\phi$. 

Interestingly, the M31 data also show an excess of aligned satellites (black circles in \Fig{phi_hist}), not unlike what we see in the Aquarius haloes. Although the number of satellites is small, a KS test returns $p_{\rm KS}=0.014$ when compared with a uniform (isotropic) $\phi$-distribution. {In an attempt to improve our statistics, we have obtained new position angle measurements from Martin et al. (2015, in preparation). These measurements are shown as green lines in \Fig{m31sats_positions}. When we include these new data (using only the most recent measurement for each satellite), the significance of the excess decreases slightly ($p_{\rm KS}=0.035$).}

Overall, these results provide suggestive (but not conclusive) evidence for significant radial alignment, suggesting that the tidal alignment mechanism discussed in the previous section might have played a role in shaping the orientation of M31 satellites. If true, this implies that the majority of M31 satellites have already completed at least one orbit around M31, placing constraints on models that ascribe the highly anisotropic spatial distribution of M31 satellites, such as the ``thin rotating plane'' of \citet{Ibata2013},  to a recent episode of accretion. On this point, we note that position angle estimates are highly uncertain for a number of recently discovered M31 satellites, including several on the aforementioned ``plane'' (see \Fig{m31sats_positions}), so the significance of this result could in principle be strengthened in the near future as new data is collected.  We are planning to extend this kind of analysis to the satellite systems of other nearby galaxies in a future contribution. 

\section{Summary and Conclusions}\label{sec:conclusions}

We have used the $N$-body simulation suite of the Aquarius Project to investigate the effect of the tidal forces exerted by a Milky Way-sized dark matter halo on the shape and orientation of its substructure. We focus on subhaloes likely to contain classical ($M_V < -8$) dwarf spheroidal galaxies according to a semi-analytic model of galaxy formation. We find that the isopotentials are very well fit by triaxial ellipsoids whose shapes can be determined accurately (to better than $5\%$) down to a convergence radius $r_{\rm conv}^{[0.1]} \approx 160$ pc for level-2 resolution Aquarius haloes. The excellent numerical resolution thus allows us to probe the dark matter potential in these systems down to the typical radii of the luminous component of dSph galaxies.

We find, in agreement with prior work, that the potential of isolated haloes is quite triaxial near the centre and becomes gradually more spherical with increasing radius. The asphericity of the potential is significant, reaching $b/a \sim 0.8$ and $c/a\sim 0.75$ near the centre, and should be taken into account when modeling the stellar kinematics of isolated dSphs. Tidal stripping, however, reduces substantially the triaxiality of subhaloes: systems that have lost more than $90\%$ of their original mass are essentially spherical. 

This result may be used to test the validity of our semi-analytic model, which predicts that luminous dSphs with relatively low dark matter content (such as Fornax and Sculptor) should be heavily stripped, and hence nearly spherical. On the other hand, faint dark matter-dominated dwarfs such as Draco and Carina are predicted to inhabit systems less affected by tides, and hence  more triaxial. Constraints on the shape of the gravitational potential in these dSphs, such as those that might be enabled by precise 3D stellar kinematics,
may thus be turned into useful constraints of their tidal evolution and of the semi-analytic model on which our predictions are based.

Regarding the orientation of subhaloes, we confirm the alignment between the major axis and the radial direction reported in earlier work and find that it holds even when considering the innermost regions of subhaloes, where the stellar component of dSph galaxies resides. Our analysis suggests that the alignment is caused by tidal torques, which tend to ``lock'' the figure rotation of a subhalo to its orbital period, so that at apocentre (where subhaloes spend most of their time) the major axis points in the radial direction.

Since the alignment takes some time to develop (it becomes significant only after second apocentric passage) its detection in a satellite population may be taken as persuasive evidence that such population has completed a number of orbits and that tidal forces have likely played an important role in its evolution. We report preliminary evidence that the satellite population of M31 is indeed radially aligned, a result that, if confirmed, would have important implications for the modeling of the assembly history of the Andromeda galaxy and its satellites. If such alignment were present in our own Galaxy, it should also have interesting implications for the modeling of the internal kinematics of the MW's satellites.

Our results seem at face value somewhat counterintuitive, for Galactic tides are often thought to stretch satellites tangentially and to enhance their asphericity. This stage, however, is short lived and confined to pericentric passages, where the tides are strongest.  The bound remnants are expected to settle into equilibrium quickly afterwards. Our analysis provides compelling evidence that, as tides operate, the remnant becomes progressively more spherical and better aligned with the radial direction. The shape and alignment of a satellite population may thus be used to gauge the importance of tidal effects on its evolution.

Our analysis can be extended and improved in several ways. One would be to identify the dynamical mechanism that leads to the sphericalization of systems affected by tides, which we present here merely as an empirical result. Another would be to explore whether satellite alignments may be used to improve the reliability of algorithms that search for primary-satellite associations, especially when applied to large datasets like the SDSS \citep[see, e.g.,][]{Wang2012,Sales2013}. Finally, it is important to investigate to what extent these results change when accounting properly for the effects of baryons, a task that must await the advent of realistic cosmological hydrodynamical simulations that can resolve the luminous regions of satellite galaxies in their proper cosmological context. All of these goals seem attainable in the near future.

\section*{Acknowledgments}
\label{sec:Acknow}

The authors are indebted to the Virgo Consortium, which was responsible for designing and running the halo simulations of the Aquarius Project. They are also grateful to Gabriella De Lucia, Amina Helmi and Yang-Shyang Li for their role in developing the semi-analytic model of galaxy formation used in this paper. They also thank Jorge Pe$\tilde{ \rm n}$arrubia and Carlos Vera-Ciro for valuable discussions and suggestions. They especially thank Nicolas Martin for generously sharing his new position angle measurements. E.S. gratefully acknowledges the Canadian Institute for Advanced Research (CIfAR) Global Scholar Academy and the Canadian Institute for Theoretical Astrophysics (CITA) National Fellowship for partial support. This work was supported in part by the National Science Foundation under Grant No. PHYS-1066293 and the hospitality of the Aspen Center for Physics.

\bibliographystyle{mn2e} 
\bibliography{paper} 

\label{lastpage}

\end{document}